\documentclass[journal]{IEEEtran}

\usepackage[dvips]{graphics}
\usepackage[dvips]{changebar}
\usepackage{amsmath}
\usepackage{rotating}
\usepackage[strings]{underscore}
\usepackage{setspace}
\usepackage{times}
\usepackage{color}
\usepackage{soul}
\usepackage{amsmath} 
\usepackage{amssymb}
\usepackage[strings]{underscore}
\usepackage{graphics,color}
\usepackage{epstopdf}
\usepackage{multicol}
\usepackage{amsmath,bm}
\usepackage{cite}
\usepackage{lipsum}
\usepackage{mathtools}
\usepackage{cuted}
\usepackage{url}

\newcommand{\mb}[1]{{\mathbf #1}}

\newcommand{\tll}{}
\newcommand{\tjj}{}
\newcommand{\tlll}{} 

\newcommand{\tl}{} 
\newcommand{\tj}{} 
\newcommand{\dc}{}

\begin{document}
\title{On the Performance of Massive MIMO Systems With Low-Resolution ADCs Over Rician Fading Channels}
\author{\IEEEauthorblockN{Tianle Liu, Jun Tong, Qinghua Guo, Jiangtao Xi, Yanguang Yu, and Zhitao Xiao}}

%

\maketitle
\newtheorem{proposition}{Proposition}

\begin{abstract}
This paper considers uplink massive multiple-input multiple-output (MIMO) systems with low-resolution analog-to-digital converters (ADCs) over Rician fading channels. Maximum-ratio-combining (MRC) and zero-forcing (ZF) receivers are considered under the assumption of perfect and imperfect channel state information (CSI). 
\dc{Low-resolution ADCs are considered for both data detection and channel estimation, and the resulting performance is analyzed.} 
Asymptotic approximations of the spectrum efficiency (SE) \dc{for large systems} are derived based on random matrix theory. With these results, we can provide insights into the trade-offs between the SE and the ADC resolution and study the influence of the Rician $K$-factors on the performance. It is shown that a large value of $K$-factors may lead to better performance and alleviate the influence of quantization noise on channel estimation. Moreover, we investigate the power scaling laws for both receivers under imperfect CSI and it shows that when the number of base station (BS) antennas is very large, without loss of SE performance, the transmission power can be scaled by the number of BS antennas for both receivers while the overall performance is limited by the resolution of ADCs. \tj{The asymptotic analysis is validated by numerical results. Besides, it is also shown that the SE gap between the two receivers is narrowed down when the $K$-factor is increased}. \tjj{We also show that ADCs with moderate resolutions lead to better energy efficiency (EE) than that with high-resolution or extremely low-resolution ADCs and using ZF receivers achieve higher EE as compared with the MRC receivers.} 
\end{abstract}


\section{Introduction}
Massive multiple-input multiple-output (MIMO)  is one of the emerging technologies for the fifth-generation  wireless communication (5G) \cite{7414384,7523234,bjornson2017massive}. Employing a large number of antennas at the base station (BS), massive MIMO systems are expected to improve the spectrum efficiency (SE) and therefore to meet the demand of high transmission rates  \cite{6736761,6457363,lu2014overview,7974726}. However, \tj{the use of massive MIMO may significantly increase the circuit power consumption, which can increase linearly with the number of  radio frequency (RF) circuits} \cite{7096297,6736746,7031971,7763844}. 
\tj{In order to} reduce the RF circuit power consumption and RF hardware \tj{cost}, the use of low-resolution analog-to-digital converters (ADCs) at the RF chains \tj{has been} widely considered \cite{7990217,7437384}.

	\tj{With} low-resolution ADCs, quantization noise arises \tj{and plays a vital role in the performance} \cite{720541}. 
	Considerable research has been conducted to understand the behavior of MIMO systems with low-resolution ADCs \cite{bai2015energy,7929310,8590717,7552476,8357527,8387969,8580585}. \tj{The \tl{achievable} spectrum efficiency (SE) with low-resolution ADCs has been analyzed using the additive-quantization-noise-model (AQNM) that provides a tractable model of the nonlinear quantization effect} \cite{7307134,7569691,7979627,JiayiZhang2016,JiayiZhang2017,7896590}. \tll{In particular}, the SE is studied in \cite{7307134} for \tll{maximum-ratio-combining (MRC)} receivers, which suggests that the SE loss due to quantization noise can be compensated by using a large number of BS antennas. 
This is further studied in \cite{7569691} when the channel state information (CSI) is \tj{obtained} by \tj{using a linear minimum mean squared error (LMMSE) estimator}. The results show \tj{that} with massive BS antennas and longer training sequences, \tj{SE comparable to that with idea ADCs can be achieved with low-resolution ADCs. 
Alternative receivers have also been considered \cite{qiao2016spectral, 7894211}. It is shown that the \tll{zero-forcing (ZF)} receiver outperforms the MRC receiver in SE when the resolution of ADCs, the numbers of users (UEs) and BS antennas are fixed. \dc{This also suggests that in order to achieve the same SE, the number of the BS antennas may be significantly reduced by using the ZF receiver, which in turn reduces the hardware costs.} 
In \cite{8063410}, the performance with mixed ADCs is analyzed and compared with that with one-bit ADCs under the assumption of a fixed total power budget. 
It is worth noting that the above studies have focused on Rayleigh fading channels.}

\tj{In practice, Rayleigh fading} may not accurately characterize the propagation environments. \tj{For applications such as} small-cell networks \cite{smallcell}, wireless-powered-internet of things (WP-IoT) \cite{8315118} and unmanned aerial vehicles (UAV) to ground transmissions \cite{8017572}, it is reasonable to consider line-of-sight (LoS) paths between the transmitters and receivers \tj{and model the channel as Rician fading}. 
In \cite{JiayiZhang2016}, Zhang \emph{et al.} investigate the SE performance of \tj{massive MIMO with low-resolution ADCs and MRC receivers over Rician fading. Asymptotic approximations of the SE for both perfect and imperfect CSI are derived} \dc{and the numerical results indicate that \tjj{SE comparable to those with ideal ADCs} can be achieved by \tl{low-resolution ADCs.}} The analysis is then extended to a mixed ADC architecture in \cite{JiayiZhang2017}. 
\tj{The power-scaling laws derived there show that the transmission power can \tl{be scaled by} the number of antennas in Rician fading channels without loss of SE, even under imperfect CSI. 
 In the aforementioned works \cite{JiayiZhang2016,JiayiZhang2017}, the CSI} is assumed to be estimated \tjj{using a small number of} ideal ADCs with a round-robin process. Specifically, at each time instant, only a subset of the BS antennas are linked with ideal ADCs and the corresponding channel states are estimated \cite{7437384}. In order to obtain the full CSI, an extended training phase is needed, which increases the overhead of training time and \tj{reduces the data transmission time \cite{7437384,8356674}.} \tj{It} may significantly reduce the training time when all the BS antennas are actively linked to low-resolution ADCs at any time. 
 \tj{This has been demonstrated in \cite{8356674} for Rayleigh fading channels, where it is shown that a system with all the low-resolution ADCs active during channel estimation may lead to a higher transmission rate compared with a mixed-ADC system with round-robin channel estimation.}

In light of the above, this paper investigates single-cell uplink MIMO systems in Rician fading channels \dc{for the scenario that only low-resolution ADCs are deployed at the receiver RF chains}. 
Assuming AQNM, we derive closed-form approximations of the SE for the MRC and ZF receivers with practical CSI. 
Using the derived approximations, we obtain the power scaling laws for both receivers.  \tjj{The simulation results show that t}he asymptotic analysis is highly accurate under different settings. It is found that the SE loss due to the use of low-resolution ADCs in the channel estimation phase can be small when the LoS path is strong, i.e., with a large value of $K$-factor, which indicates the feasibility of systems employing only low-resolution ADCs  in such applications. 
The ZF receivers generally outperform the MRC receivers in SE with a realistic number of BS antennas but their gap can be narrowed when the $K$-factors are increased. This is 
because when the LoS path becomes \tl{strong}, the inter-UE-interference \tjj{mainly arise from} UEs with similar angles of arrival (AoA).  
Moreover, when the number of BS antennas goes infinite, the non-LoS part in channel fading for UEs tends to be orthogonal, which is favorable \tj{in terms of interference minimization}. In this asymptotic case, the power-scaling laws indicate the same SE \tj{for both receivers}. Furthermore, the transmission power can be scaled down by the number of BS antennas for a given SE while the overall SE is limited by the resolution of ADCs. \tjj{The energy efficiency (EE) of the system is also evaluated under Rician fading by using a system-level power consumption model. The numerical results show that the EE can be improved by employing ADCs with moderate resolutions. Similar to the SE performance, the EE is higher when the LoS paths become stronger and the ZF receiver is used.}
  
\tj{The paper is organized as follows. We introduce the system model 
in Section II. The uplink SE is analyzed for the MRC and ZF receivers under perfect CSI and the asymptotic approximations for ZF are provided in Section III. The SE and corresponding asymptotic \dc{analysis} are studied for the case with imperfect CSI in Section IV. The analysis is 
validated with numerical results in Section V. The conclusions are given in Section VI.} 

\section{System Model}
\label{sm}
Consider the uplink of a single-cell MIMO system with $M$ BS antennas and $K$ single-antenna UEs. The cell is circular with radius $d_{\rm max}$. Let $H_{m k}$ be the $(m, k)$-th entry of $\mb H$, denoting the channel gain between the $k$-th UE and the $m$-th BS antenna. 
We assume $H_{mk} =  \sqrt{ \beta_{k}} \widetilde{H}_{mk}$, where $\beta_{k}$ represents the large scale fading and $\widetilde{H}_{mk} $ characterizes the small scale fading between UE-$k$ and the $m$-th BS antenna
\begin{equation}
\label{system1}
\mb H  =   \widetilde{\mb H} \mathrm{diag}(  \sqrt{ \beta_{1}},  \sqrt{ \beta_{2}},  \cdots,  \sqrt{ \beta_{K} }).    
\end{equation}
We consider Rician fading channels which consist of $\bar{\mb H}$ due to LoS propagation and 
$\mb H_w$ due to non-LoS propagation. Specifically, $\widetilde{\mb H}$ between the BS and UEs can be modeled as 
\begin{equation}
\label{fastfading}
\widetilde{\mb H}= \bar{\mb H}\left[ \mb{\Omega}\left(\mb{\Omega}+\mb I_K \right)^{-1}  \right]^{1/2} + {\mb H}_w \left[ \left(\mb{\Omega}+\mb I_K \right) ^{-1} \right]^{1/2}
\end{equation}
where $\mb{\Omega}$ is a $K \times K$ diagonal matrix with $ \Omega_{k,k} = \mathcal{K}_k$ and $\mathcal{K}_k$ is the $K$-factor for UE-$k$, which determines the ratio between the power gains of the LoS component and non-LoS component. 
\dc{Following \cite{JiayiZhang2016,JiayiZhang2017}, we assume a uniform linear array (ULA) with half-wavelength spacing, i.e., $d=\lambda/2 $, where $d$ is the antenna spacing and $\lambda$ the wavelength}. 
The LoS component 
\begin{equation}\label{LosH2}
\bar{ H}_{mk}=e^{-j\left(m-1 \right)\pi  \sin\left( \theta_k \right)  } 
\end{equation} 
where $\theta_k$ is the \tl{AoA} for UE-$k$ \dc{uniformly distributed in $\left[-\pi/2, \pi/2 \right]$.} 
 \tl{The channel response between UE-$k$ and the $m$-th antenna over Rician fading can be modeled as a Gaussian distribution with a non-zero mean, i.e.,
 \[ 
 {\widetilde{H}_{mk}}  \sim \mathcal{CN} \left(\sqrt{\frac{\mathcal{K}_k \tl{\beta_{k}}}{\mathcal{K}_k+1}}  \bar{ H}_{mk}, \frac{\beta_k}{\mathcal{K}_k+1} \right). 
 \]}
\vspace{-0.04cm}
Assuming equal transmission power for all UEs, the observed signals at the BS is 
\begin{equation}
	\label{MUMIMO}
	\mb y = \sqrt{P_u}\mb H   \mb x  + \mb n = \sqrt{P_u}\sum_{k=1}^{K} \mb h_k   x_k + \mb n, 		 
\end{equation}
where $\mb y \in  \mathbb{C}^{M}$ is the observed signal vector at the BS, $\mb H = [\mb h_1, \mb h_2, ... ,\mb h_K ] \in \mathbb{C}^{M \times K}$ is the channel matrix, 
$\mb x \in \mathbb{C}^{K }$ contains the transmitted \dc{unit-power} symbols $\{x_k\}$ from the $K$ UEs, $P_{u}$ is the transmission power for all UEs, and $\mb n \in \mathbb{C}^{M}$ is the additive white Gaussian noise (AWGN) with variance $\sigma^{2}$ (in Joule/symbol). 

At the BS, the received analog signals are sampled and quantized into digital signals with finite-resolution ADCs. The quantized signal $\mb y_q$ can be approximated by the AQNM, which decomposes the quantizer output as the summation of the attenuated input signal and an uncorrelated distortion, i.e.,  
\begin{equation}
	\label{adcsystem}
	\mb y_{q} = \alpha \mb y + \mb n_q = \alpha \sqrt{P_u}\mb H \mb x + \alpha \mb n + \mb n_q, 
\end{equation}
where $\mb n_{q}$ is the additive Gaussian quantization noise which is uncorrelated with $\mb y$. The coefficient $\alpha$ can be obtained by $\alpha= 1- \epsilon $, \tj{where $\epsilon$ is the inverse of the single-to-quantization-noise ratio ($\rm SQNR$)}. Given the number of quantization bits, $\epsilon$ can be found in \cite{7307134}. 
The covariance matrix of  $\mb n_{q}$ is  
\begin{equation}
	\label{rqq}
	\mb R_{\mb n_{q}} =\mathbb{E} \left[ \mb n_{q} \mb n_{q}^{H} | \mb H \right]= 
	\alpha \left(1-\alpha\right) {\rm diag} \left(P_u \mb H \mb H^{H} + \sigma^{2} \mb I_M  \right).
\end{equation}
\tjj{This is to approximate the quantization noise at different receiver \dc{RF chains} uncorrelated, similarly to \cite{7307134,7569691,7979627,JiayiZhang2016,JiayiZhang2017,7896590}.} In this paper, \tlll{following \cite{7307134,7569691,7979627,JiayiZhang2016,JiayiZhang2017,7896590}, we further} assume that $\mb x$, $\mb n$ and $\mb n_q$ are Gaussian-distributed and mutually independent. In addition, each entry in  $\mb x$, $\mb n$ and $\mb n_q$ is with zero means and variances given by $1$, $\sigma^2$ and $\mb R_{\mb n_q}$, respectively.
\section{SE With Perfect CSI }
 At the BS, linear receiver filters are employed to estimate the transmitted signals for recovering data.
Assuming perfect CSI, the estimate of UE-$k$'s signal is given as  
\begin{equation}
\label{adcout}
\hat{x}_{k} = \mb{g}_{k}^{H}\mb{y}_q,  
\end{equation} 
where $\mb{g}_{k}$ is the receiver filter for UE-$k$. Let  $\gamma_k$ denote the \tlll{signal-to-interference-plus-noise ratio (SINR)} at the BS. The uplink gross SE $\bar{R}_{k}$ for UE-$k$ is  
\begin{equation}
\label{rate}
\bar{R}_{k} = \log \left(1+\gamma_k \right).  
\end{equation}
We next discuss the ${\gamma}_k$ for MRC and ZF with perfect CSI.
\subsection{SE Analysis}
\subsubsection{MRC Receiver}
\dc{For the MRC receiver, the filter for UE-$k$} is given by  
 \begin{equation} \label{MRCfilter}
	\mb g_k = [\mb H ]_{:,k}   . 
	\end{equation} 
The estimated signal for UE-$k$ is written as \begin{equation} \label{xkhat}
	 \hat{x}_{k} = \alpha \sqrt{P_u} || \mb {h}_{k}||^2  x_k + \alpha \sqrt{P_u} \tl{\sum\limits_{n=1, n \neq k }^K} \mb {h}_{k}^{H} \tl{\mb {h}_{n}  x_n}  + \alpha \mb{h}_{k}^{H} \mb n +   \mb{h}_{k}^{H} \mb{n}_q. 
	 	 \end{equation} 
	The SINR $\gamma_k$ is computed as 
	\begin{equation}\label{SNRmrc}
	\gamma_k = \frac{\alpha^2 P_u \| \mb h_k \| ^4}{{\Theta}_{k}},  
	\end{equation}
	where ${\Theta}_{k}$ denotes the distortion-plus-noise power: 	
	\begin{equation} \label{NI}
	{\Theta}_{k}=     \alpha^2 \| \mb h_k \| ^2 \sigma^2 +  \mb h_k ^H \mb{R_{n_q}} \mb h_k + \alpha^2 P_u \tl{\sum\limits_{n=1, n \neq k }^K} |\mb h_k^H \tl{\mb h_n}|^2.  
	\end{equation} 
 The first term on the right hand side (RHS) of (\ref{NI}) denotes the channel noise power while the second term stands for the ADC quantization noise power and the last term is power of the inter-UE interference.

	\subsubsection{ZF Receiver}
	ZF receivers generally outperform MRC in terms of interference suppression. When the ZF receivers are used, the filter for UE-$k$ is 
      \begin{equation} \label{zffilter}
	\mb g_k = [\mb H \left(\mb H ^{H} \mb H \right)^{-1}]_{:,k}.    
	\end{equation} 
 It is \dc{known that the ZF receiver is able to fully mitigate the inter-UE interference. Thus} the estimated signal in (\ref{xkhat}) can be further modeled by  
	\begin{equation}
	\hat{x}_{k} = \alpha \sqrt{P_u}  x_k + \alpha \mb{g}_{k}^{H} \mb n +   \mb{g}_{k}^{H} \mb{n}_q, 
	\end{equation} 
and the SINR is 
\begin{equation}\label{SNRzf}
\gamma_k = \frac{\alpha^{2}  P_u}{{\Theta}_{k}},  
\end{equation}
where ${\Theta}_{k}$ denotes the distortion-plus-noise power and \tl{is} given as
\begin{equation}\label{DPNzf}
{\Theta}_{k} = \alpha^2 \|\mb{g}_{k}\|^2 \sigma^2 +   \  \mb{g}_{k}^{H}\mb{R_{n_q}} \mb{g}_{k}. 
\end{equation} 

\subsection{Asymptotic Analysis}
\label{app}
In this subsection, we discuss the asymptotic analysis of the SE using random matrix theory under the assumption of large systems. We refer to \tl{\cite{JiayiZhang2016} for the} treatment for the MRC receiver and focus on the derivation for the ZF receivers in Rician fading channels. 

From (\ref{DPNzf}), the distortion-plus-noise power ${\Theta}_{k}$ consists of two parts, i.e., channel noise power $\alpha^2 \|\mb{g}_{k}\|^2 \sigma^2$ and the ADC quantization noise power $\mb{g}_{k}^{H} \mb R_{\mb n_{q}} \mb{g}_{k}$, \tl{where
\begin{equation}
\|\mb{g}_{k}\|^2 = \left[ \left( \mb H^H \mb H\right)^{-1}  \right]_{k,k} \approx \mathbb{E} \left[ \left( \mb H^H \mb H\right)^{-1}  \right]_{k,k},
\end{equation}\tlll{ where the expectation is w.r.t. the small-scale fading.} 
From \cite[Theorem 4]{QiZhang2014}, $\mb H^H \mb H$ follows a non-central Wishart distribution and the squared Euclidean norm of UE-$k$'s filter satisfies   
\begin{equation}\label{appfilterzf}
\|\mb{g}_{k}\|^2 \tlll{\approx}\frac{\left[  {\mb \Sigma}^{-1} \right]_{k,k} }{\beta_k\left( M-K\right) }, 
\end{equation} 
where 
 \begin{equation}\begin{split}
  {\mb \Sigma}\triangleq&\left(\mb{\Omega}+\mb I_K \right)^{-1} \\&+\frac{1}{M} \left[ \mb{\Omega}\left(\mb{\Omega}+\mb I_K \right)^{-1}  \right]^{1/2} \bar{\mb H}^H \bar{\mb H}\left[ \mb{\Omega}\left(\mb{\Omega}+\mb I_K \right)^{-1}  \right]^{1/2}.
 \end{split}
 \end{equation}} 
For the quantization noise, the following approximation \dc{tends to be accurate in a large system}, 
\begin{equation} \label{approxqn}  
 \mb{g}_{k}^{H} \mb R_{\mb n_{q}} \mb{g}_{k} \approx \mb{g}_{k}^{H} \mathbb{E} \left[ \mb R_{\mb n_{q}}\right]  \mb{g}_{k}, 
\end{equation} 
where \dc{the expectation is w.r.t. the small-scale fading}. 
From (\ref{fastfading}) and (\ref{LosH2}),   
\begin{equation}
\label{HHkk} 
 \dc{\tll{\mathbb{E}}[|\widetilde{H}_{m,k}|^2]=} \frac{\mathcal{K}_k}{\mathcal{K}_k+1}+\frac{1}{\mathcal{K}_k+1} =1.  
\end{equation}
With (\ref{HHkk}), we can approximate (\ref{rqq}) by
  \begin{equation} \label{ERnq}
\mathbb{E} \left[ \mb R_{\mb n_{q}}\right] \approx \alpha \left(1-\alpha\right) \left( P_u \sum\limits_{k=1}^K \beta_k+\sigma^2 \right) \mb I_M. 
\end{equation}  
Substituting  (\ref{appfilterzf}) and (\ref{ERnq}) into (\ref{approxqn}), the power of the quantization noise can be approximated as   
\begin{equation}
\label{adcNOISEZF}
\mb{g}_{k}^{H} \mb R_{\mb n_{q}} \mb{g}_{k} \approx \alpha \left(1-\alpha\right) \left( P_u \sum\limits_{k=1}^K \beta_k+\sigma^2 \right) \frac{\left[ \tl{ {\mb \Sigma}^{-1}} \right]_{k,k} }{\beta_k\left( M-K\right) }.
\end{equation}
Finally, substituting  (\ref{appfilterzf}) and (\ref{adcNOISEZF}) into  (\ref{DPNzf}), we can find the distortion-plus-noise power for the ZF receiver and approximate the SINR in (\ref{SNRzf}) by  
\begin{equation} 
\begin{split}
\gamma_k \approx \frac{ P_u\beta_k\left( M-K\right)}{\left(\frac{1}{\alpha} \sigma^2 +  \left(\frac{1}{\alpha}-1\right)  P_u \sum\limits_{\tl{n=1}}^K \tl{\beta_n} \right) \left[ \tl{{\mb \Sigma}^{-1}} \right]_{k,k}  }. 
\end{split}
\end{equation}
This can be used to predict the SE performance along with (\ref{rate}).
\section{SE With Imperfect CSI}
 In practice, the CSI is often obtained during a training phase. In this paper, we consider the LMMSE channel estimation. 
Following \cite{JiayiZhang2016,JiayiZhang2017,QiZhang2014,5340650} we assume that the large-scale fading coefficient, the deterministic component $\bar{\mb{H}}$, and $\mb \Omega$ are known at the BS.  
\subsection{LMMSE Channel Estimation}
 \label{LMMSECE}
 \tl{In the training phase, each UE
 transmits a pilot signal of length $L$ to the BS and the pilot signals for all UEs can be denoted by a $K \times L$ matrix $\mb \Phi$, where 
 $\mb \Phi   =  \left[  \bm \phi_1,  \bm \phi_2, \cdots,  \bm \phi_{K} \right]^{T} \in \mathbb{C}^{K\times \dc{L}} $. The received signals, $ {\mb Y}^{t} \in  \mathbb{C}^{ M \times L}$, at the BS are given by  
 \begin{equation}
 {\mb Y}^{t} = \sqrt{P_t}\mb H \mb\Phi+\mb N^{t}, 
 \end{equation}
and the quantized received signals can be modeled by AQNM as 
\begin{equation}
	\label{system}
	{\mb Y}^{t}_q = \alpha \dc{\sqrt{P_t}} \mb H \mb \Phi +   \alpha \mb N^{t} + \mb N_{q}^{t},
\end{equation} where \tl{$P_t$ is the pilot transmission power which is assumed to be the same as $P_{u}$}} in the data transmission phase for all the UEs. The channel noise and ADC quantization noise are denoted by $\mb N^{t}$ and $\mb N_{q}^t$, respectively. \tl{Let $ \widetilde{\mb{H}}_w \triangleq \mb{H}_w {\rm diag} \left\lbrace   \sqrt{\beta_1}, \cdots, \sqrt{\beta_K} \right\rbrace$ and \dc{assume orthogonal training signals with $ {\mb \Phi} \mb \Phi^H  \triangleq L \mb I_K$} \cite{QiZhang2014}.} After removing the LoS part which is assumed known at the BS,  (\ref{system})  can be simplified to 
\dc{\begin{eqnarray}
     \label{simplesystem}
     \tl{{\mb Y}^{t}_q} \!\!\!\!&=&\!\!\!\!  \alpha \sqrt{ P_u} \widetilde{\mb{H}}_w \left[ \left( \mb \Omega + \mb I_K\right)^{-1} \right]^{1/2} \mb \Phi +   \alpha \mb N^{t} + \mb N_{q}^{t}. 
\end{eqnarray}
}
For the \tl{$l$-th pilot symbol},  
\tl{\begin{equation}
\label{ytpilot}
{{\mb y}^{t}_{q}}_l = \alpha \sqrt{P_u} \sum_{k=1}^{K}\widetilde{\mb{h}}_{w,k}  \dc{\phi_{k,l}} \left( \mathcal{K}_k+1\right)^{-1/2} + \alpha \mb n^{t}_l + \mb n_{q,l}^{t}.
\end{equation}
and}
\begin{equation}
\tl{\mb R_{\mb n_{q}}^t}= \alpha \left(1-\alpha\right) {\rm diag}\left( \mb R_{\mb y^{t}_{l}}\right) 
\end{equation} 
where 
\begin{equation}
\tl{\mb R_{\mb y^{t}_{l}} =  \mathbb{E} \left[ \mb y^{t}_{l} \mb y^{t H}_{l} \right]}   
\end{equation}
   and 
   \begin{equation}
   \tl{\mb R_{\mb n_{q} }^t} =    \alpha\left( 1-\alpha\right)  \left(  \sigma^2  +    P_{u}\sum_{n=1}^{K} \beta_{n} \right) \mb I_M.
   \end{equation}

\dc{Now let us consider the LMMSE estimation of UE-$k$'s channel $\mb h_k$.} 
Let $\mb x_k$ denote the $k$-th row of $\mb X$ where $\mb X\triangleq \sqrt{P_u}\mb \Phi$ and $\widetilde{\mb y}^{t}_{k}  \triangleq  \mb Y^{t}_q \mb x_k^H$. We have 
\begin{equation}
\begin{split}
 \widetilde{\mb y}^{t}_{k}   & =\tl{ \alpha \widetilde{\mb{H}}_w \left[ \left( \mb \Omega + \mb I_K\right)^{-1} \right]^{1/2}\mb X \mb x_k^H  +   \alpha \mb N^{t} \mb x_k^H+ \mb N_{q}^{t} \mb x_k^H } \\& 
= \alpha L \tl{P_u 
\widetilde{\mb{h}}_{w,k}\left( \mathcal{K}_k+ 1\right)^{-1/2}} + \alpha  \mb N^{t} \mb x_k^H+\mb N_{q}^{t} \mb x_k^H. 
\end{split}
	\label{tildey}
\end{equation} 
For each UE-$k$, assume that the channel vector $\widetilde{\mb{h}}_{w,k}$ follows i.i.d. zero-mean Gaussian distribution with variance $\beta_k$. The LMMSE estimate of $\widetilde{\mb{h}}_{w,k}$ is then computed as 
\begin{equation} \label{htildek}
\widehat{\mb h}_{w,k}  = {\psi}_k  \widetilde{\mb y}^{t}_{k}, 
\end{equation}
where 
\begin{equation}
\label{psikone}
\psi_k  =  \mb C_{\widetilde{\mb{h}}_{w,k} , \widetilde{\mb y}^{t}_{k} } \mb R_{\widetilde{\mb y}^{t}_{k}   } ^{-1} 
\end{equation}
and \tl{$ \mb C_{\widetilde{\mb{h}}_{w,k} , \widetilde{\mb y}^{t}_{k} }$} is the covariance matrix of \tl{$\widetilde{\mb{h}}_{w,k}$} and $\widetilde{\mb y}^{t}_{k}$, which can be calculated by 
\begin{equation}
\mb C_{\widetilde{\mb{h}}_{w,k} , \widetilde{\mb y}^{t}_{k} } = \alpha  P_u L \beta_k \tl{\left( \mathcal{K}_k+ 1\right)^{-1/2}}\mb I_M.
\end{equation}
 $\mb R_{\widetilde{\mb y}^{t}_{k} }$ is the variance matrix of $\widetilde{\mb y}^{t}_{k}$ and 
 \begin{equation}\begin{split} 
 \mb R_{\widetilde{\mb y}^{t}_{k} }  = & \alpha^2 \dc{L}^2\beta_k  P_{u}^2\tl{\left( \mathcal{K}_k+ 1\right)^{-1}} \mb I_M + \alpha^2 \sigma^2 L   P_{u} \mb I_M \\ 
&+  L P_{u}  \alpha\left( 1-\alpha\right)  \left(  \sigma^2  +  P_u \sum_{n=1}^{K} \beta_{n}      \right) \mb I_M  .
 \end{split}\end{equation}
 \tj{Now the channel estimator for UE-$k$ in (\ref{psikone}) can be written as}  
 \begin{equation}\begin{split}
  \psi_k &= \tl{\frac{\beta_{k}\left( \mathcal{K}_k+ 1\right)^{-1/2}}{\alpha L \beta_k  P_{u}  \left( \mathcal{K}_k+ 1\right)^{-1} +  \sigma^2 + \left( 1-\alpha\right)    P_u \sum\limits_{n=1}^{K} \beta_{n}   }  }.
 \end{split}
\end{equation}
Recall that the deterministic component $\bar{\mb{H}}$ and $\mb \Omega$ are known at the BS. 
The channel estimation error for UE-$k$ 
\[
\tj{{\mb e_k} \triangleq  {\mb h_k} - \widehat{\mb h}_k= \frac{1}{\sqrt{ \mathcal{K}_k+1 }} \left(  \mb h_{w,k}   - \widehat{\mb h}_{w,k} \right)}.   
\]
\dc{It can be shown that the} variance of the entries in $\mb e_k$ can be computed as 
\begin{equation}
	\sigma_{{e}_{k}}^2 =      \frac{\tl{\beta_{k} \left( 1-\xi_k \right)}}{ \mathcal{K}_k+1} 
\end{equation}
where 
\begin{equation}\label{xi}\begin{split}
\xi_k  = \frac{P_u L\beta_{k} }{P_u L \beta_{k}  +  \tl{ \left( \mathcal{K}_k+ 1\right)\left[ \frac{\sigma^2}{\alpha} +  \left( \frac{1}{\alpha}-1\right)  P_u  \sum\limits_{n=1}^K \beta_n  \right] }}. 
\end{split}
\end{equation}
Correspondingly, the variance of the estimate of the non-LoS component of the channel for UE-$k$ is 
 \begin{equation}\label{CEnonLos}
 \widehat{\sigma}_k^2 =  \frac{\beta_{k} \xi_k }{\mathcal{K}_k+1}. 
 \end{equation}\vspace{-0.8cm}
\subsection{ SE Analysis} 
The data transmission phase follows the training phase. Let $\widehat{\gamma}_k$ denote the SINR with imperfect CSI. The uplink SE is given by  
\begin{equation}
\label{imrate}
\bar{R}_{k} = \log \left(1+ \widehat{\gamma}_k \right). 
\end{equation}
 We now discuss $\widehat{\gamma}_k$ for the MRC and ZF receivers with imperfect CSI.
\subsubsection{MRC Receiver}
When the MRC receiver is used, the filter for UE-$k$ 
\begin{equation} \label{immrcfilter}
\widehat{\mb g}_k = [\widehat{\mb H} ]_{:,k}.    
\end{equation} 
The SINR $\widehat{\gamma}_k$ is given by 
\begin{equation}\label{imSNRmrc}
\widehat{\gamma}_k = \frac{\alpha^2 P_u \| \widehat{\mb h}_k \| ^4}{{\widehat{\Theta}}_{k}},  
\end{equation}
where $\widehat{\Theta}_{k}$ denotes the distortion-plus-noise power 
\tl{\begin{equation} \label{imNIMRC}\begin{split}
\widehat{\Theta}_{k}=&\alpha^2 \| \widehat{ \mb h}_k \| ^2 \sigma^2 +\widehat{\mb h}_k ^H \mb{R_{n_q}} \widehat{\mb h}_k + \alpha^2 P_u \sum\limits_{n=1, n \neq k }^K |\widehat{\mb h}_k^H \widehat{\mb h}_n|^2 \\&+\alpha^2 P_u \sum\limits_{n=1}^K \|\widehat{\mb h}_k \|^2\sigma_{{e}_{n}}^2,
\end{split}
\end{equation} which} consists of contributions from the channel noise, ADC quantization noise, inter-UE interference and channel estimation error. 
\subsubsection{ZF Receiver}
When the ZF receiver is used, \begin{equation} \label{imzffilter}
\widehat{\mb g}_k = [\widehat{\mb H} \left(\widehat{\mb H} ^{H} \widehat{\mb H} \right)^{-1}]_{:,k}.    
\end{equation} 
Similar to the case with perfect CSI, the SINR  $\widehat{\gamma}_k$ with ZF receivers is given as
\begin{equation}\label{imSNRzf}
\widehat{\gamma}_k = \frac{\alpha^2 P_u }{{\widehat{\Theta}}_{k}},  
\end{equation}
where 
\begin{equation} \label{imNIzf}
\widehat{\Theta}_{k}=     \alpha^2 \| \widehat{ \mb g}_k \| ^2 \sigma^2 +  \widehat{\mb g}_k ^H \mb{R_{n_q}} \widehat{\mb g}_k +\alpha^2 P_u \tl{\sum\limits_{n=1}^K} ||\widehat{\mb g}_k ||^2 \tl{\sigma_{{e}_{n}}^2}.
\end{equation} 
\tl{In particular, the covariance matrix of quantization noise, $\mb{R_{n_q}}$, in (\ref{imNIMRC}) and (\ref{imNIzf}) is defined in (\ref{rqq}).} \dc{Note also that in contrast to the case with perfect CSI, the ZF receiver cannot fully mitigate the inter-UE interference, as seen from the last item of (\ref{imNIzf}).} 
\subsection{Asymptotic Analysis} 
\label{imcsiasy}
We next derive asymptotic approximations of the SINR for the MRC and ZF receivers with imperfect CSI, which can provide insights about the influence of the $K$-factor, ADC resolution, and number of BS antennas on the SE. From the analysis in Section \ref{LMMSECE}, \dc{the estimated Rayleigh fading channel vectors for different UEs $\widehat{\mb h}_{w}$ are mutually independent and the entries can be modeled as i.i.d. random variables (RVs) for a given UE. For UE-$k$, the entries of $\widehat{\mb h}_{w,k}$ can be modeled by }  
\[
   \widehat{H}_{w, mk} \sim \mathcal{CN} (0, \beta_{k} \tl{\xi_k}),  
\] 
where \tl{$\xi_k$} is given in (\ref{xi}). Let 
\begin{equation}
\label{Hmk}
    \widehat{H}_{mk} = \sqrt{\frac{\mathcal{K}_k \beta_{k}}{\mathcal{K}_k+1}}\rho_{mk} + \sqrt{\frac{1}{\mathcal{K}_k+1}} \delta_{mk},  
\end{equation} 
where from (\ref{LosH2}) 
\[
    \rho_{mk} \triangleq  e^{-j\left(m-1 \right)\pi  \sin\left( \theta_k \right)  } = \rho_{mk}^c-j\rho_{mk}^s 
\]  
and 
\[ 
\delta_{mk} \triangleq  \delta_{mk}^c + j \delta_{mk}^s
\]
with zero mean and variance of $\frac{\beta_{k} \tl{\xi_k}}{2}$ \tl{for independent real and imaginary parts $\delta_{mk}^c$ and $\delta_{mk}^s$}. 
In the following, we assume large systems and derive asymptotic approximations of the SE.  

\subsubsection{MRC Receiver}
 In order to find the asymptotic approximation to (\ref{imSNRmrc}) for the MRC receiver, the approximation expressions of $\| \widehat{\mb h}_k \| ^2, \| \widehat{\mb h}_k \| ^4$ and $|\widehat{\mb h}_k ^H \widehat{\mb h}_n|^2$ are needed. 
\tj{As derived in the Appendix, \tl{when} $M \rightarrow \infty$,}
 \begin{equation}\label{hh}
		\| \widehat{\mb h}_k \| ^2  \approx \frac{M \beta_{k}\left( \mathcal{K}_k + \tl{\xi_k} \right) }{\mathcal{K}_k+1},
	\end{equation} 
	\begin{equation}\label{h^4} \begin{split}
		\| \widehat{\mb h}_k \| ^4  \approx \frac{M\beta_{k}^2\left( 2\mathcal{K}_k \tl{\xi_k} + 2 M  \mathcal{K}_k \tl{\xi_k} + M   \mathcal{K}_k  ^2 + \left(M + 1 \right) \tl{\xi_k^2}\right)}{\left( \mathcal{K}_k+1\right) ^2}, \end{split}
	\end{equation} 
		and  \begin{equation}\label{hh^2}
		\begin{split}
		| \widehat{\mb h}_k^H \widehat{\mb h}_n| ^2 \approx \frac{\beta_{n}\beta_{k} \left( \mathcal{K}_k  \mathcal{K}_n  \lambda_{kn}^2+M \mathcal{K}_k\tl{\xi_n}+M \mathcal{K}_n\tl{\xi_k}+M \tl{\xi_n\xi_k}\right) }{\left( \mathcal{K}_k+1\right)\left( \mathcal{K}_n+1\right)}.
		\end{split}
	\end{equation} 
The \tlll{SINR} $\widehat{\gamma}_k$ of the system for the MRC receiver can then be approximated by 
\begin{equation}\label{imappSNRmrc}
\widehat{\gamma}_{k} \approx \frac{ \tl{\alpha^2} P_u \frac{M\beta_{k}^2}{\left( \mathcal{K}_k+1\right) ^2} \left( 2\mathcal{K}_k\tl{\xi_k} + 2 M  \mathcal{K}_k \tl{\xi_k} + M   \mathcal{K}_k  ^2 + \left(M + 1 \right) \tl{\xi_k^2}\right)}{{\widehat{\Theta}}_{k}},  
\end{equation}where $\widehat{\Theta}_{k}$ is given as
\begin{equation} \begin{split}\label{imappNIMRC}
\widehat{\Theta}_{k}  &\triangleq \alpha^2 \|  \widehat{ \mb h}_k \| ^2 \sigma^2 +  \widehat{\mb h}_k ^H \mb{R_{n_q}} \widehat{\mb h}_k +  \alpha^2 P_u\sum\limits_{n=1, n \neq k }^K |\widehat{\mb h}_k^H \tl{\widehat{\mb h}_n|^2} \\& \;\; + \alpha^2 P_u \tl{\sum\limits_{n=1}^K} \|\widehat{\mb h}_k^H \|^2\sigma_{{e}_{\tl{n}}}^2 \\
& \approx \frac{M \beta_{k}\bigg(\mathcal{K}_k + \tl{\xi_k}\bigg)}{\mathcal{K}_k+1}\left(\alpha\sigma^2+\alpha\left( 1-\alpha\right) P_u\sum\limits_{k=1}^K\beta_k+\alpha^2 P_u \sum\limits_{n=1}^K \sigma_{{e}_{n}}^2   \right) \\& \; \;+ \alpha^2 P_u\sum\limits_{n=1, n \neq k }^K \frac{\beta_{n}\beta_{k}}{\left( \mathcal{K}_k+1\right)\left( \mathcal{K}_n+1\right)}( \mathcal{K}_k  \mathcal{K}_n  \lambda_{kn}^2+M \mathcal{K}_k\tl{\xi_n}\\& \;\; +M \mathcal{K}_n\tl{\xi_k}+M \tl{\xi_n\xi_k}).
\end{split}
\end{equation}

\subsubsection{ZF Receiver}
It is known that when $M$ is large, for the ZF filter for UE-$k$,
\begin{equation}
\label{imzfg2}
\tl{\|\widehat{\mb{g}}_{k}\|^2 = \left[ \left( \widehat{\mb H}^H \widehat{\mb H}\right)^{-1}  \right]_{k,k} \approx \mathbb{E} \left[ \left( \widehat{\mb H}^H \widehat{\mb H}\right)^{-1}  \right]_{k,k}.} 
\end{equation} 
\dc{Similarly to the case with perfect CSI,} \tl{$ \widehat{\mb H}^H \widehat{\mb H}$} follows a non-central Wishart distribution.  Using a similar strategy as \cite[(53), (54)]{QiZhang2014}, we can obtain  
\begin{equation}\label{imappfilterzf}
\|\widehat{\mb{g}}_{k}\|^2 
\approx \frac{\left[ \widehat {\mb \Sigma}^{-1} \right]_{k,k} }{ \tl{\beta_{k}}\left( M-K\right)  }, 
\end{equation} 
where 
\begin{equation}\begin{split}
\label{sigmahat}
\tl{\widehat {\mb \Sigma}}&\triangleq \tl{\mb \Xi} \left(\mb{\Omega}+\mb I_K \right)^{-1} \\& \quad +\frac{1}{M} \left[ \mb{\Omega}\left(\mb{\Omega}+\mb I_K \right)^{-1}  \right]^{1/2} \bar{\mb H}^H \bar{\mb H}\left[ \mb{\Omega}\left(\mb{\Omega}+\mb I_K \right)^{-1}  \right]^{1/2}, 
\end{split}
\end{equation} 
\tl{and $\mb \Xi$ is a $K \times K$ diagonal matrix with $\Xi_{k,k}=\xi_k$, where $\xi_k$ is defined in (\ref{xi}). }
 \dc{As such, the SINR} 
 \begin{equation}\label{imSINRzf}
 \dc{\widehat{\gamma}_k = \frac{ \alpha^2 P_u }{{\widehat{\Theta}}_{k}}}
 \end{equation}
for the ZF receivers can be approximated by using 
\begin{equation}
\label{imNIzfapp}
\tl{\widehat{\Theta}_{k}   \approx  \frac{   \left[ \alpha \sigma^2 +    \alpha \left( 1-\alpha\right) P_u \sum\limits_{n=1}^K \beta_n + \alpha^2 P_u \sum\limits_{i=1}^K \sigma_{{e}_{i}}^2 \right]   \left[ \widehat {\mb \Sigma}^{-1} \right]_{k,k}}{\beta_{k} \left( M-K \right) }}.
\end{equation} 

\subsection{\tl{Power-Scaling Laws and Influence of ADC Resolution}}
\subsubsection{MRC Receivers}
The asymptotic approximations derived above allow the study of the power scaling law for MRC receivers with imperfect CSI, which provides insight on the performance when the number of antennas increases. 
Consider the scenario $M \to \infty$. \tl{Multiply $\frac{1}{M^2}$ to the numerator and denominator of (\ref{imSNRmrc}) and let $P_u=\frac{E_u}{M^{\nu}}$, where  $\nu  > 0 $ and $E_u$ is a fixed value.} We have  
\begin{equation}\label{nulling1}
\alpha^2 \frac{E_u}{M^{\nu+2}} |\widehat{\mb h}_k^H \widehat{\mb h}_i|^2= \alpha^2 \frac{E_u}{M^{\nu}} | \frac{1}{M}\widehat{\mb h}_k^H \widehat{\mb h}_i|^2 \overset{a.s.}{\to} 0.
\end{equation}
  Following (\ref{nulling1}), removing the relevant zero items in (\ref{imSNRmrc}), it then reduces to 

\tl{\begin{equation}\label{reduceimSNRmrc}
\widehat{\gamma}_k = \frac{ \frac{E_u}{M^{\nu}} \| \widehat{\mb h}_k \| ^2}{\frac{\sigma^2}{\alpha}   + \frac{E_u}{M^{\nu}}  \sum\limits_{n=1}^K \beta_n \left( \frac{1-\alpha}{\alpha} +\frac{1-\xi_n}{\mathcal{K}_n+1} \right) }. 
\end{equation}
Considering $\frac{\alpha E_u}{\sigma^2M^{\nu}}\|\widehat{\mb h}_k\|^2=\frac{\alpha E_u}{\sigma^2M^{\nu-1}}|\frac{1}{M}\dc{\widehat{\mb h}_k^H}\widehat{\mb h}_k|$ and following the approximation of $\| \widehat{\mb h}_k\|^2$ in (\ref{hh}), the SINR tends to be}
\begin{equation}
\begin{split}
\widehat{\gamma}_k 
&\tj{\overset{M \rightarrow \infty}{\rightarrow}}  \frac{\alpha E_u}{\sigma^2M^{\nu-1}} \left | \frac{1}{M}\widehat{\mb h}_k^H\widehat{\mb h}_k \right|\\&\tl{\approx}\frac{\alpha E_u}{\sigma^2 M^{\nu-1}}\left ( \frac{\mathcal{K}_k\beta_k}{ \mathcal{K}_k+1 }+\frac{\tl{\xi_k}\beta_k}{\mathcal{K}_k+1}  \right)\\&=\frac{\alpha \mathcal{K}_k\beta_k E_u}{\sigma^2 M^{\nu-1}\left(\mathcal{K}_k+1\right) }   +  \\& \quad \frac{\alpha E_u^2 \beta_k^2L}{\sigma^2 M^{2\nu-1}\left( \mathcal{K}_k+1\right) \left(L\frac{E_u}{M^{\nu}} + \frac{\sigma^2}{\alpha} + (\frac{1}{\alpha}-1)\frac{E_u}{M^{\nu}}\sum\limits_{n=1}^K \beta_n \right) }.  
\end{split}
\end{equation}
It is easy to observe that when $M \to \infty$, the SINR 
\begin{equation} \label{gammakmrcminf}
\widehat{\gamma}_k \tll{\to} \frac{\alpha \mathcal{K}_k\beta_k E_u}{\sigma^2 M^{\nu-1}\left(\mathcal{K}_k+1\right) }   +   \frac{\alpha E_u^2 \beta_k^2 L}{M^{2\nu-1}\left( \mathcal{K}_k+1\right) \frac{\sigma^4}{\alpha} }.
\end{equation}
The power scaling law for the MRC receiver can now be discussed from (\ref{gammakmrcminf}).

When the system operates in Rayleigh fading channels, i.e., $\mathcal{K}_k=0$, the SINR tends to be a constant value 
\begin{equation} \label{gammakmrcrayminf}
\widehat{\gamma}_{k, \rm mrc} \tll{\overset{M \rightarrow \infty}{\rightarrow}} \frac{\alpha^2 E_u^2 \beta_k^2 L}{ \sigma^4 } 
\end{equation}
if $\nu=0.5$. 
This shows that under Rayleigh fading, the transmission power can be scaled by the square root of the number of BS antennas. 
\tj{From (\ref{gammakmrcrayminf}), when low-resolution ADCs are used for both the channel estimation and data transmission phases, the SINR decreases quadratically with $\alpha$ which characterizes the quantization error. This differs from \cite[(24)]{JiayiZhang2016}, which shows the SINR decreases linearly with $\alpha$ when ideal ADCs  are used (in a round-robin manner) for the LMMSE channel estimation.}

For Rician fading channels, i.e., $\mathcal{K}_k \ne 0$, the SINR tends to be a constant value 
\begin{equation}
\widehat{\gamma}_{k, \rm mrc} \tll{\overset{M \rightarrow \infty}{\rightarrow}} \frac{\alpha \mathcal{K}_k\beta_k E_u}{\sigma^2 \left(\mathcal{K}_k+1\right) } 
\end{equation}
if $\nu=1$.   
This suggests that in Rician fading channels, the transmission power can be scaled linearly by the number of BS antennas when the SE is fixed. 
In contrast to the case of Rayleigh fading, the SE loss caused by the usage of low-resolution ADCs for channel estimation diminishes when $M\to\infty$, and the asymptotic SINR becomes the same as \cite[(25)]{JiayiZhang2016} which assumes ideal ADCs for channel estimation.

\subsubsection{ZF Receivers} 

We next derive the power scaling law for the ZF receivers under imperfect CSI, following the same treatments as for the MRC receivers. 
Assuming $M \to \infty$ and removing the zero items when $M \to \infty$, (\ref{imSNRzf}) then becomes   
\begin{equation}\label{imSNRzfm}
         \widehat{\gamma}_k \tll{\overset{M \rightarrow \infty}{\rightarrow}} \frac{ E_u }{\frac{1}{\alpha} M^{\nu}\| \widehat{ \mb g}_k \| ^2 \sigma^2 }. 
\end{equation}

From \cite[Corollary 5]{QiZhang2014}, when $M$ is large, $\frac{1}{M}\bar{\mb H}^H \bar{\mb H}$ \tl{can be approximated by} an identity matrix. \tjj{Following this, substituting (\ref{imappfilterzf}) into
  (\ref{imSNRzfm}) leads to } 
\begin{equation}\begin{split}
&\quad \frac{ E_u }{\frac{1}{\alpha} M^{\nu}\| \widehat{ \mb g}_k \| ^2 \sigma^2 } \\&\tll{\approx} \frac{ E_u \left(M-K \right)  }{\frac{1}{\alpha} M^{\nu}\sigma^2    } \left\lbrace    \frac{\mathcal{K}_k\beta_k}{ \mathcal{K}_k+1 }   +   \frac{\tl{\xi_k}\beta_k}{\mathcal{K}_k+1}  \right\rbrace \\&=\frac{ \alpha E_u \left(1-\frac{K}{M} \right)  }{ M^{\nu-1}\sigma^2    } \left\lbrace    \frac{\mathcal{K}_k\beta_k}{ \mathcal{K}_k+1 }   + \frac{\tl{\xi_k}\beta_k}{\mathcal{K}_k+1}  \right\rbrace\\&=\frac{\alpha E_u \mathcal{K}_k\beta_k}{ \left( \mathcal{K}_k+1\right) M^{\nu-1}\sigma^2  }   + \\&  \frac{ \alpha E_u \beta_{k} \frac{E_u}{M^{\nu}}         L  \beta_k }{\left( \mathcal{K}_k+1\right) M^{\nu-1}\sigma^2\left(  \frac{E_u}{M^{\nu}}         L  \beta_k + \frac{\sigma^2}{\alpha} + \left( \frac{1}{\alpha } -1 \right) \frac{E_u}{M^{\nu}} \sum\limits_{n=1}^K \beta_{n}  \right)  } 
\\&=\frac{\alpha E_u \mathcal{K}_k\beta_k}{ \left( \mathcal{K}_k+1\right) M^{\nu-1}\sigma^2  }+\frac{ L \alpha^2 E_u^2 \beta_k^2 }{\left( \mathcal{K}_k+1\right) M^{2\nu-1}\sigma^4 }.  
\end{split}
\end{equation}

For Rayleigh fading channels, i.e., $\mathcal{K}_k= 0$, the SINR tends to be
\begin{equation}
\widehat{\gamma}_{k, \rm zf} \tll{\overset{M \rightarrow \infty}{\rightarrow}} \frac{\alpha^2 E_u^2 \beta_k^2  L}{\sigma^4 }  
\end{equation}
when $\nu= 0.5$. 
For Rician fading channels, i.e., $\mathcal{K}_k\ne 0$, the SINR tends to be
\begin{equation}
\widehat{\gamma}_{k, \rm zf} \tll{\overset{M \rightarrow \infty}{\rightarrow}}\frac{\alpha \mathcal{K}_k\beta_k E_u}{\sigma^2 \left(\mathcal{K}_k+1\right) } 
\end{equation}
when $\nu=1$. 
Therefore, for a given SE, the transmission power can be scaled by $M$ for Rician fading and by $\sqrt{M}$ for Rayleigh fading, where $M$ is the number of BS antennas. 
It is observed that the ZF and MRC receivers have the same asymptotic expressions of the SINR when $M \to \infty$. This is reasonable as using a large number of antennas the MRC and ZF receivers exhibit   similar performance due to channel hardening \cite{CEhard} where the channels of different UEs become orthogonal to each other.

\tj{It can be observed from the above asymptotic analysis that, for both receivers, the SE loss due to the use of low-resolution ADCs in channel estimation may be lower if strong LoS paths are present, which encourages the usage of low-resolution ADCs in some practical scenarios such as small cells and UAV transmissions. In fact, it can be verified from (\ref{CEnonLos}) that as the $K$-factor increases, the channel estimation error caused by ADC quantization noise diminishes  and $\widehat{\gamma}_k$ improves. 
 }   
 
\subsection{\tj{Very Strong LoS Paths}} 
We next consider scenarios where the channels have very strong LoS paths. For simplicity, assume that all UEs have the same value of $K$-factors, i.e., $\mathcal{K}_i=\mathcal{K}_k, \forall i$, and $\mathcal{K}_k\to \infty$. 
For the MRC receiver, \tl{(\ref{imappSNRmrc})} tends to 
\begin{equation}
\widehat{\gamma}_k \tll{\to} \frac{P_u M^2 \beta_{k}}{M \left( \frac{\sigma^2}{\alpha} + \tl{\frac{1-\alpha}{\alpha}} P_u \sum\limits_{n=1 }^K \beta_{n}\right)+  P_u \sum\limits_{n=1, n \neq k }^K \beta_{n}\lambda_{kn}^2 }. 
\end{equation}
This indicates that when the LoS paths become strong, only the UEs with similar AoAs lead to inter-UE-interference. It is also observed that the overall SE performance is limited by the resolution of ADCs. 

For the ZF receiver, as $\mathcal{K}_k\to \infty$, \tl{(\ref{imSINRzf}) tends to} 
\begin{equation}\label{knzfim}
\dc{\widehat{\gamma}_k \tll{\to} \frac{\beta_{k} P_u \left( M-K\right) }{ \left[   \frac{\sigma^2}{\alpha}  +    \left( \frac{1-\alpha}{\alpha} \right)  P_u \sum\limits_{n=1}^N \beta_n  \right]\tl{\left[ \left( \frac{1}{M} \bar{\mb H}^H \bar{\mb H}\right) ^{-1} \right]_{k,k}} } }.
\end{equation}
\tll{Similar to the MRC receiver, the overall SE is limited by the resolution of ADCs.}

\section{Numerical Results }

We now present numerical results to validate the analysis in Section IV. We consider a single-cell, uplink system with $K=10$ UEs. The large scale fading coefficients $\{\beta_{k}\}$ are set the same as \cite{JiayiZhang2017}. The $K$-factors for all UEs are assumed the same, following \cite{JiayiZhang2017}, \cite{QiZhang2014}. The same low-resolution ADCs are adopted for the channel estimation and data transmission phases. The pilot length for each UE is fixed to $L=K$ except for Fig. \ref{pilot}. \tj{The AoAs of different UEs are generated randomly and the average SE is measured.} 

\begin{figure}[!t]
	\centering
	\includegraphics[ width=3.4in]{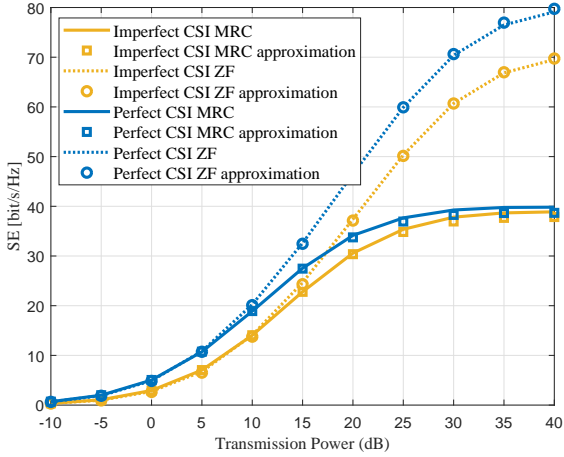}
	\caption{Uplink SE versus transmission power, with 
		$M = 200, K = 10$, $3$-bits ADCs and Rician $K$-factor of 0 dB. \tl{The MRC approximation with perfect CSI follows \cite{JiayiZhang2016}.}} 
	\label{powerK0db}
\end{figure}

\subsection{SE Performance}
Fig. \ref{powerK0db} shows the SE performance under different values of transmission power, where $M=200$ BS antennas equipped with 3-bits ADCs are assumed. The simulation results are compared with the approximation results derived in Section \ref{app} and \ref{imcsiasy} for both the MRC and ZF receivers. It is observed that the approximation results are highly accurate. 
The ZF receivers show  similar SE performance to the MRC receivers when the system operates with a relatively low transmission power while it generally outperforms MRC with a high transmission power.
This is because at high SNR the inter-UE-interference dominates the distortion-plus-noise power.
For ZF receivers, the inter-UE-interference is assumed perfectly canceled and the gap between the cases with perfect and imperfect CSI is due to the channel estimation error. The SE for the MRC receivers with imperfect CSI and perfect CSI are very close when the transmission power is very high. This implies that the inter-UE-interference plays a more significant role in the distortion-plus-noise power as compared to the channel estimation error when the transmission power increases.  
\begin{figure}[t]
	\centering
	\includegraphics[ width=3.4in]{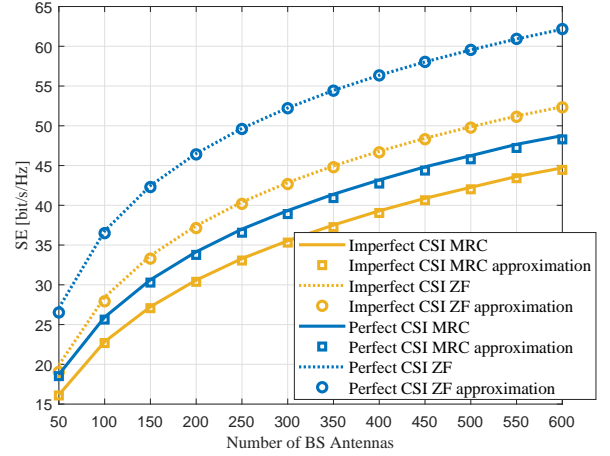}
	\caption{ Uplink SE versus number of BS antennas with $K = 10, P_u = 20$ dB, $3$-bits ADCs, and Rician $K$-factor  of 0 dB. }
	\label{antennaM}
\end{figure}

Fig. \ref {antennaM} evaluates the SE performance of the two receivers with different number of antennas. Again, the approximation results are highly accurate. It is shown that the increased number of antennas improves the SE performance for both receivers. This is because with such a great number of antennas the non-LoS channels become orthogonal to each other. \tj{As such,
only the UEs with similar AoAs could contribute to inter-UE-interference and the MRC receiver suffers less inter-UE-interference. }
\begin{figure}[t]
	\centering
	\includegraphics[ width=3.4in]{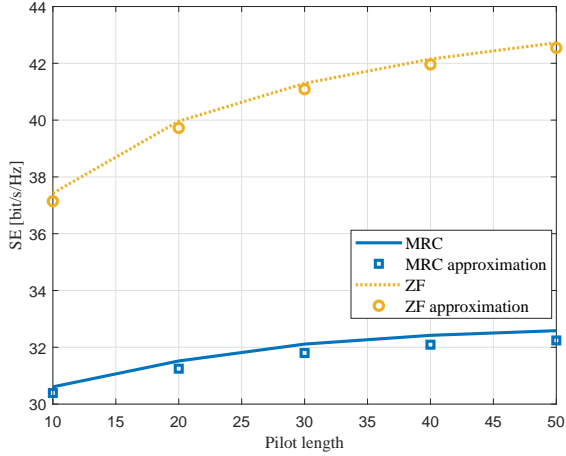}
	\caption{Uplink SE versus pilot length, with
		$M = 200, K = 10, P_u = 20$ dB, $3$-bits ADCs and Rician $K$-factor of 0 dB.  }
	\label{pilot}
\end{figure}

Fig. \ref{pilot} shows the influence of the pilot length $L$. It is seen that the gross SE increases with $L$ for both receivers as channel estimation error decreases. This, however, is achieved at the cost of  increased training overheads, which may in turn affect the net SE due to the reduction of the effective data transmission time when the coherence interval is fixed.  
\begin{figure}[t]
	\centering
	\includegraphics[ width=3.4in]{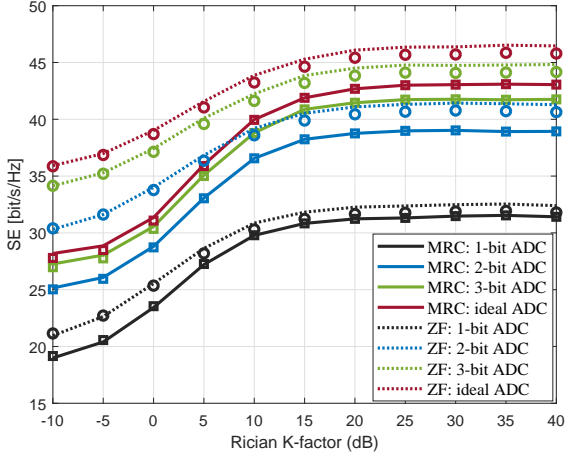}
	\caption{Uplink SE versus ADC resolution and Rician $K$-factor, with
		$M = 200, K = 10, P_u = 20$ dB. The solid and dash lines stand for the simulation results of MRC and ZF respectively while the square and circle marks represent the asymptotic results for MRC and ZF.}
	\label{Kn}
\end{figure}

Fig. \ref{Kn} presents the SE for different ADC resolutions and $K$-factors with CSI obtained from LMMSE channel estimation. It is observed that the SE increases with $K$-factor, benefiting from the reduction of channel uncertainty when the $K$-factor increases. This is consistent with the asymptotic analysis in Section \ref{imcsiasy}. To achieve a similar SE, the requirement of the ADC resolution can be alleviated when the $K$-factor is larger. It is also seen that with 3-bits ADCs the SE for both receivers are already close to that with idea ADCs, confirming the feasibility of using low-resolution ADCs for both channel estimation and data transmission in Rician fading channels. The improvement of SE is more significant with the MRC receiver as the $K$-factor increases, as stronger LoS paths not only reduce the channel uncertainty but also mitigate the inter-UE-interference. 
This also leads to the observation that the gap in the SEs of the ZF and MRC receivers is narrower when the $K$-factors become large. 
The loss of using 1-bit ADCs tends to be significant due to the excessive quantization noise, even when the $K$ factor is large. 
\begin{figure}[t]
	\centering
	\includegraphics[ width=3.4in]{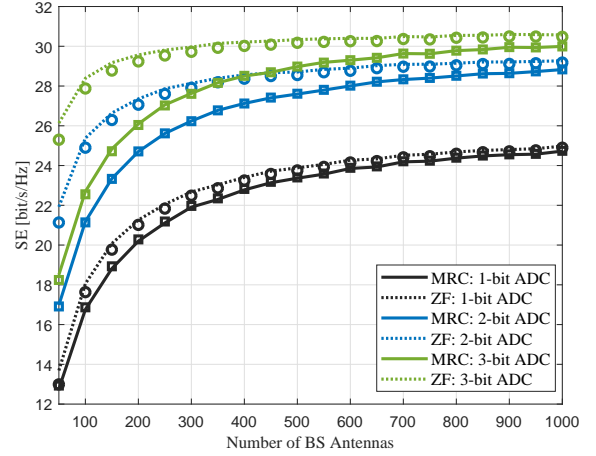}
	\caption{Demonstration of the power scaling laws with $K = 10, E_u = 40$ dB and $K$-factor of 0 dB. The square and circle marks represent the asymptotic results for MRC and ZF, which are compared with the simulation results of MRC and ZF in solid and dash lines respectively.}
	\label{scalepu}
\end{figure}

Fig. \ref{scalepu} finally verifies the power scaling laws discussed in Section \ref{imcsiasy}. The transmission power is scaled by the number of BS antennas. When the number of antennas $M$ increases to a very large value, with the same ADCs the SE for the MRC and ZF receivers tend to be constants and show very similar values. This verifies our prediction in the asymptotic analysis, i.e., the transmission power can be reduced by a factor of $M$ in Rician fading channels and the SE of the two receivers tend to be close when $M \to \infty$. In addition, the SE is limited by the ADC resolution. 
The ZF receiver converges with fewer antennas than the MRC receiver. This suggests  that in practical applications, the ZF receiver may be preferred if a smaller number of BS antennas is desired.

\tll{\subsection{EE Performance}}
We also consider the EE performance measured by the number of bits transmitted per Joule: 
\begin{equation} \label{EEE}
{\rm EE}=  \frac{\sum_{k=1}^K R_k}{\rm P}, 
\end{equation}
where $R_k$ denotes the effective transmission rate of UE-$k$ and  $\rm P$ is the average power consumption. Following \cite{tianletwc}, we assume coherence blocks of $U=1800$ symbols and  uplink ratio $\zeta^{\rm ul}=0.4$. Thus, $U\zeta^{\rm ul}=720$ symbols are transmitted in the uplink in each coherence block. Given the pilot length $L$ and the bandwidth $B=20$ MHz, the effective uplink transmission rate of UE-$k$ is  
\begin{equation}
\label{achievable rate}   
R_k = \zeta^{\rm ul} \left(1- \frac{ L }{ U  \zeta^{\rm ul}} \right) B \bar{R}_k,  
\end{equation} 
where $1- \frac{L}{ U \zeta^{\rm ul}}$ characterizes the training overhead. The average power consumption  
\begin{equation}
\label{c}
\begin{split}
{\rm P} =& M {\rm \left( \rm P_{BS} + 2\rm P_{ADC} \right)}  +K\rm P_{UE}+\rm P_{SYN}\\&+\rm P_{other}+\rm P_{CD} + \rm P_{CE} +\rm P_{SD},
\end{split}
\end{equation}
where $\rm P_{BS}, \rm P_{UE}, \rm P_{SYN}, \rm P_{CD}$ are the RF circuit power consumption per RF chain at the BS, the total circuit power consumption of each UE's device, the power of the local oscillator at the BS, the power for channel coding and decoding respectively. $\rm P_{other} $ in (\ref{c}) represents the backhaul power consumption and the power of site cooling, etc. The values of $\rm P_{BS}, \rm P_{UE}, \rm P_{SYN}, \rm P_{CD}$ and $\rm P_{other}$ are set the same as in \cite{tianletwc}. The power consumption of a $b$-bits ADC is given by 
\begin{equation}
\label{ADCpowerold}
{\rm P_{\rm ADC}}= {\rm FOM} \cdot f_{s} \cdot 2^{b}, 
\end{equation}
where the figure of merit (FOM) means energy consumed per conversion step and $ f_{s} $ the sampling rate. In the following simulations, we set the value of $\rm FOM $ the same as in \cite{tianletwc} and consider Nyquist sampling rate to model commercialized ADCs. 
The digital signal processing (DSP) power consumption in the channel estimation is
\begin{equation}
{\rm	P_{CE} }=  \frac{B }{U}  \frac{{\rm C_{ CE}}}{L_{\rm BS}},  
\end{equation}
where $L_{\rm BS}$ represents the computational efficiency at the BS and is assumed as $L_{\rm BS}=12.8 $ Gflops/W.
The computational complexity for estimating the $K$ UEs' channels is approximately ${\rm C_{ CE}}=	2LKM+2K^2+MK$ flops. We also take into account the power consumed by signal detection: 
\begin{equation}
{\rm P_{SD}} = \ B \zeta^{\rm ul}  \left(1- \frac{\ K \tau^{\rm ul} }{ \zeta^{\rm ul}   U}   \right) \frac{\rm C_{symbol}}{ L_{BS}}+ \rm P_{\rm BL},  
\end{equation}
where $\rm C_{\rm symbol}$ is the complexity of recovering a symbol:
\begin{equation}
{\rm C_{\rm symbol}}= 2KM-K \; {\rm flops}
\end{equation} for both receivers. 
The computational power consumed due to the filter calculation, ${\rm P_{BL}}$, is given as  
	\begin{equation}
{\rm P_{BL}} = \frac{\ B  {\rm C_{wm}}}{ U L_{\rm BS}},
\end{equation}
where ${\rm C_{wm}}$ is the complexity involved in obtaining the filters. Specifically, for ZF receivers ${\rm C_{wm}}=\frac{1}{3}K^{3}+3K^{2}M+KM-\frac{1}{3}K\; {\rm 	flops}$. For MRC receivers, as shown in (\ref{MRCfilter}) and (\ref{immrcfilter}), the filters are directly obtained from the channel matrix. Thus, the complexity for finding the MRC filter is ignored. 

\begin{figure}[t]
	\centering
	\includegraphics[ width=3.4in]{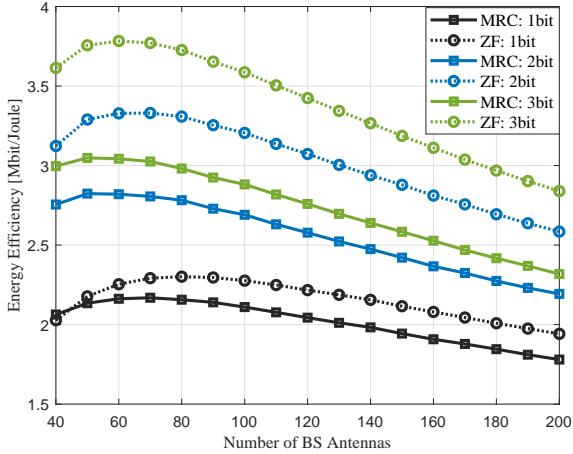}
	\caption{Simulation results for uplink EE versus the number of antennas with
		$K = 10, P_u = 20$ dB and Rician K-factor $= 0$ dB. The solid lines stand for the simulation results of MRC receivers while the dash lines represent the EE performance with ZF receivers.   }
	\label{EEM}
\end{figure}

\begin{figure}[t]
	\centering
	\includegraphics[ width=3.4in]{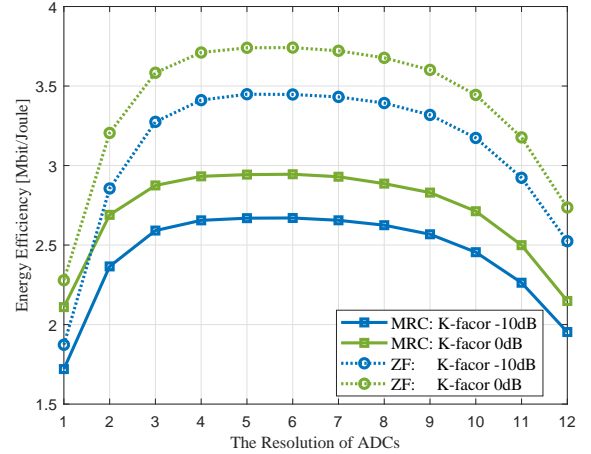}
	\caption{Simulation results for uplink EE versus ADC resolution and Rician $K$-factor, with
		$M = 100, K = 10, P_u = 20$ dB. The blue lines stand for the simulation results of MRC and ZF with Rician K-factor = -10 dB while the green lines represent the EE performance with Rician K-factor = 0 dB.   }
	\label{EEKn}
\end{figure}

 Fig. \ref{EEM} shows the achievable EE of the system with different numbers of BS antennas over Rician fading. It is shown that the EE is not monotone with the number of antennas and can be lower with a large number of antennas, e.g., $M=200$, than that with a moderate number of antennas, e.g., $M=80$. This is due to the significant RF circuit power consumption for a large system.  

 Fig. \ref{EEKn} shows the achievable EE with different resolutions of ADCs as well as different values of the Rician $K$-factors for the MRC and ZF receivers. The EE performance is higher for both receivers when the LoS paths are stronger (i.e., with a larger Rician-$K$ factor). Moreover, ADCs with a moderate resolution, e.g., 5-bits, achieve higher EE than that with high-resolution or extremely low-resolution ADCs. It is also observed that the ZF receivers generally outperform the MRC receivers.  This indicates that although MRC receivers require lower computational complexity than  ZF receivers, the reduction of the power consumption due to signal processing can not compensate for the loss in the \tll{SE}.   
\vspace{0.5cm} 

\section{Conclusion }\vspace{-0.5cm}
This paper investigates MIMO systems with low-resolution ADCs operating over Rician fading channels. We study the potential of using low-resolution ADCs in both the channel estimation and data transmission phases. We derive the asymptotic approximations of the SINR for the MRC and ZF receivers under imperfect CSI. The numerical results demonstrate that the derived approximations are highly accurate for different numbers of antennas, ADC resolutions and Rician $K$-factor. The feasibility of using low-resolution ADCs for acquiring the CSI in Rician fading channels is demonstrated. The analysis and simulations show the channel estimation error caused by using low-resolution ADCs is alleviated when strong LoS paths are present. For a very large number of BS antennas, the SE of the two receivers tend to be close and are limited by the ADC resolution. However, the ZF receivers generally perform better in SE when a moderate number of BS antennas is used. \tll{The numerical results also indicate that a higher EE can be achieved by using ADCs with moderate resolutions. Besides, stronger LoS paths and using ZF receivers are able to further increase the EE. }

\appendices
\section{}
\label{FirstAppendix}
In order to find the asymptotic approximation to (\ref{imSNRmrc}) for the MRC receiver, let us analyze $\| \widehat{\mb h}_k \| ^2, \| \widehat{\mb h}_k \| ^4$ and $|\widehat{\mb h}_k ^H \widehat{\mb h}_n|^2$. 

We start with $\| \widehat{\mb h}_k \| ^2$. As $M \rightarrow \infty$, 
\[ 
\| \widehat{\mb h}_k \| ^2 \approx \sum\limits_{m=1}^{M} \mathbb{E} \left( | \widehat{H}_{mk}|^2 \right). 
\]
From (\ref{Hmk}), 
\begin{equation} 
\begin{split}
\| \widehat{\mb h}_k \| ^2 & \approx\\ & 
\frac{ \sum\limits_{m=1}^{M}\mathbb{E}  \left(  \mathcal{K}_k \beta_{k} |\rho_{mk}|^2 + 2 \sqrt{\mathcal{K}_k\beta_{k}} \mathcal{R}(\rho_{mk}^* \delta_{mk}) +  |\delta_{mk}|^2 \right)}{\mathcal{K}_k+1}.
\end{split}
\end{equation}
Noticing that \tj{$|\rho_{mk}|^2=1$, $\mathbb{E}(|\delta_{mk}|^2)=\beta_k \tl{\xi_k} $,} and $\mathbb{E}(\delta_{mk})=0$, we have  
\begin{equation}\label{h^2}
	\| \widehat{\mb h}_k \| ^2  \approx \frac{1}{\mathcal{K}_k+1} \sum\limits_{m=1}^M \mathbb{E} \left(  \mathcal{K}_k\beta_k +     \beta_{k} \tl{\xi_k}    \right) = \frac{M \beta_{k}\left( \mathcal{K}_k + \tl{\xi_k} \right) }{\mathcal{K}_k+1}.
\end{equation} 
We next study 
\dc{\begin{equation}\begin{split}
	\| \widehat{\mb h}_k \| ^4 
	& \approx \bigg[  \sum\limits_{m=1}^{M}\mathbb{E} \left(  |\widehat{  H}_{mk}|^2\right) \bigg]  ^2 \\&
	= \frac{1}{\left( \mathcal{K}_k+1\right) ^2}\bigg\{ \sum\limits_{m=1}^{M}\mathbb{E}  \bigg[  \mathcal{K}_k \beta_{k} |\rho_{mk}|^2 \\&\quad  +\sqrt{\mathcal{K}_k\beta_{k}}(\rho_{mk}^* \delta_{mk} + \rho_{mk} \delta_{mk}^*) 
	+|\delta_{mk}|^2 \bigg]\bigg\} ^2\\& 
	=
	\frac{1}{\left( \mathcal{K}_k+1\right) ^2}\mathbb{E}   \bigg[   \left( \sum\limits_{m=1}^{M} \mathcal{K}_k \beta_{k} |\rho_{mk}|^2  \right)^{2} \\&\quad +
	\left( \sum\limits_{m=1}^{M} \mathcal{K}_k \beta_{k} \rho_{mk}^{*}
	\delta_{mk} \right)^{2} + \left( \sum\limits_{m=1}^{M} \mathcal{K}_k \beta_{k} \rho_{mk}\delta_{mk}^{*} \right)^{2}  \\&\quad +  \bigg( \sum\limits_{m=1}^{M}|\delta_{mk}|^2  \bigg)^{2}  +   2   \sum\limits_{m=1}^{M}\mathcal{K}_k\beta_{k}\left(\rho_{mk}^{*} \delta_{mk} \right)^{2} \\&\quad +
	2  \sum\limits_{m_1=1}^{M}\sum\limits_{m_2=1}^{M} \mathcal{K}_k\beta_{k}   |\rho_{m_1k}|^2   |\delta_{m_{2,k}}|^2       \bigg]
	\end{split}\end{equation}
}
Invoking the assumptions in (\ref{Hmk}) and after some mathematical manipulations  
we have   
\begin{equation}
\begin{split}
	&\| \widehat{\mb h}_k \| ^4 \approx \\& \frac{2M\mathcal{K}_k\beta_{k}  \widehat{\beta}_{k} + 2 M^2  \mathcal{K}_k \beta_{k} \widehat{\beta}_{k} + \left( M \mathcal{K}_k \beta_{k}\right) ^2 + \left(M^2 + M \right) \widehat{\beta}_{k}^2 }{\left( \mathcal{K}_k+1\right) ^2.}\end{split}
\end{equation} 
\dc{After some simplifications, we find (\ref{h^4}).} 

We finally consider 
\begin{equation}\begin{split}
\label{imkn2}
| \widehat{\mb h}_k^H \widehat{\mb h}_n| ^2 &\approx  \left| \sum\limits_{m=1}^{M}\mathbb{E}   \left(\widehat{  H}_{mk}^{*} \widehat{  H}_{mn}   \right)\right|^2 \\&=\frac{1}{\left( \mathcal{K}_k+1\right)\left( \mathcal{K}_n+1\right) }\bigg|\bigg( \sum\limits_{m=1}^{M}\mathbb{E}   \sqrt{\mathcal{K}_k \beta_{k}\mathcal{K}_n \beta_{n}}\rho_{mk}^* \rho_{mn}  \\&\quad + \sqrt{\mathcal{K}_k\beta_{k}} \rho_{mk}^* \delta_{mn} + \sqrt{\mathcal{K}_n\beta_{n}} \rho_{mn} \delta_{mk}^* + \delta_{mk}^*\delta_{mn}\bigg) \bigg|^2
\end{split}\end{equation}

\dc{Following the analysis in Section \ref{imcsiasy}, $\delta_{mk}$ is independent of $\delta_{mn}$ for $\forall n \ne k$ and $\mathbb{E}(\delta_{mk})=0$. Removing the zero items, the remaining items in (\ref{imkn2}) are given as} 
\begin{equation}\begin{split}\label{imkn3}
| \widehat{\mb h}_k^H \widehat{\mb h}_n| ^2 & \tll{\approx} \frac{1}{\left( \mathcal{K}_k+1\right)\left( \mathcal{K}_n+1\right)}\mathbb{E}\bigg[  \left( \sum\limits_{m=1}^{M} \sqrt{\mathcal{K}_k \beta_{k} \mathcal{K}_n \beta_{n}} | \rho_{mk}^* \rho_{mn}|\right) ^2 \\&\quad +\left( \sum\limits_{m=1}^{M}\sqrt{\mathcal{K}_k\beta_{k}} |\rho_{mk}^* \delta_{mn}|\right)^2\\&\quad +\left( \sum\limits_{m=1}^{M} \sqrt{\mathcal{K}_n\beta_{n}} |\rho_{mn} \delta_{mk}^*|\right)^2 + \left( \sum\limits_{m=1}^{M}|\delta_{mk}^*\delta_{mn}| \right) ^2  \bigg]
\end{split}\end{equation}
We now analyze the four terms on the RHS of \tl{(\ref{imkn3})} in the bracket individually. 
Define 
\[
\lambda_{kn} \triangleq \frac{{\rm sin} \left( \frac{M\pi}{2} \left( {\rm sin}\theta_k -{\rm sin} \theta_n\right)  \right)    }{ {\rm sin}\left( \frac{\pi}{2} \left( {\rm sin}\theta_k -{\rm sin} \theta_n\right)  \right)    }.
\]
Following  \cite[(118)]{QiZhang2014}, the first term leads to 
\tl{\begin{equation}\label{imLoSapp}
\begin{split}
&\mathbb{E}\left( \sum\limits_{m=1}^{M} \sqrt{\mathcal{K}_k \beta_{k} \mathcal{K}_n \beta_{n}}  |\rho_{mk}^* \rho_{mn}|\right) ^2 \\&
= \mathcal{K}_k \beta_{k} \mathcal{K}_n \beta_{n}  \bigg\{  \lambda_{kn}^2  {\rm cos}^2 \left[  \frac{M-1}{2}\pi \left( {\rm sin}\theta_{k} - {\rm sin} \theta_{n}\right)    \right] \\& \quad +  \lambda_{kn}^2  {\rm sin}^2 \left[  \frac{M-1}{2}\pi \left( {\rm sin}\theta_{k} - {\rm sin} \theta_{n}\right)    \right] \bigg\} \\&
= \mathcal{K}_k \beta_{k} \mathcal{K}_n \beta_{n} \lambda_{kn}^2.
\end{split}\end{equation}}
The second part can be approximated by 
\begin{equation}\begin{split}
& \mathbb{E} \left( \sum\limits_{m=1}^{M}\sqrt{\mathcal{K}_k\beta_{k}} |\rho_{mk}^* \delta_{mn}|\right) ^2 \\
&= \mathcal{K}_k \beta_k \mathbb{E} \bigg[ \sum\limits_{m=1}^M \left(  \rho_{mk}^{*}\delta_{mk} \right)^2 \\&\quad + \sum\limits_{m_1=1}^{M}\sum\limits_{m_2=1,m_2\neq m_1}^{M} \left(  \rho_{m_{1,k}}^{*}\delta_{m_{1,k}} \right) \left(  \rho_{m_{2,k}}^{*}\delta_{m_{2,k}} \right) \bigg] \\&
=\mathcal{K}_k \beta_k  \mathbb{E}  \bigg[ \sum\limits_{m=1}^M \left( \rho_{mk}^c\delta_{mn}^c-\rho_{mk}^s\delta_{mn}^s \right)^2  \\&\quad +   \sum\limits_{m=1}^M \left( \rho_{mk}^s\delta_{mn}^c+\rho_{mk}^c\delta_{mn}^s \right)^2  \bigg] \\&= M \mathcal{K}_k\beta_{k} \beta_{n}\tl{\xi_n}. 
\end{split}\end{equation}
Following a similar process, the third term is approximated by 
\begin{equation}
\mathbb{E} \left[ \sum\limits_{m=1}^{M}\sqrt{\mathcal{K}_n\beta_{n}} \rho_{mn}\delta_{mk}^* \right] ^2= M \mathcal{K}_n\beta_{n} \beta_{k}\tl{\xi_k}.
\end{equation}
The last term is approximated by 
\begin{equation}\begin{split}
&\mathbb{E} \left( \sum\limits_{m=1}^{M} \delta_{mk}^* \delta_{mn}\right) ^2 =  \bigg[ \sum\limits_{m=1}^M \left(  \delta_{mk}^{*}\delta_{mn} \right)^2 \\&+ \sum\limits_{m_1=1}^{M}\sum\limits_{m_2=1,m_2\neq m_1}^{M} \left(  \delta_{m_1,k}^{*}\delta_{m_1,n} \right) \left(  \delta_{m_2,k}^{*}\delta_{m_2,n} \right) \bigg]
\end{split}
\label{immrc4}
\end{equation}
Removing the vanishing terms, (\ref{immrc4}) simplifies to 
\begin{equation}\begin{split}
&\mathbb{E} \left( \sum\limits_{m=1}^{M} \delta_{mk}^* \delta_{mn}\right) ^2 \\&= \mathbb{E}  \bigg[ \sum\limits_{m=1}^M\left( \delta_{mk}^{c}\delta_{mn}^c + \delta_{mk}^{s}\delta_{mn}^s\right)^2 + \sum\limits_{m=1}^M\left( \delta_{mk}^{c}\delta_{mn}^s - \delta_{mk}^{s}\delta_{mn}^c\right)^2\bigg] \\&= M \beta_{n}\beta_{k}\tl{\xi_n\xi_k}.
\end{split}\end{equation}
Finally,  
\begin{equation}
	\begin{split}
	&| \widehat{\mb h}_k^H \widehat{\mb h}_n| ^2\approx\\&\frac{ \mathcal{K}_k \beta_{k} \mathcal{K}_n \beta_{n} \lambda_{kn}^2+M \mathcal{K}_k\beta_{k} \beta_{n}\tl{\xi_n}+M \mathcal{K}_n\beta_{n} \beta_{k}\tl{\xi_k}+M \beta_{n}\beta_{k}\tl{\xi_n\xi_k}}{\left( \mathcal{K}_k+1\right)\left( \mathcal{K}_n+1\right)}.\end{split}
\end{equation}
\tl{The expression in (\ref{hh^2}) can then be \dc{obtained} by some simple algebraic operations.}
\medskip
\bibliographystyle{ieeetr}
\bibliography{Preprint}
\end{document}